\documentclass[journal,12pt,onecolumn]{IEEEtran}

\usepackage{amsfonts,color,cite}
\usepackage{amssymb}
\usepackage{algorithmic}
\usepackage{amsmath}
\usepackage{algorithm}
\usepackage{cases}
\floatname{algorithm}{Algorithm}

\usepackage{multirow} %
\usepackage{booktabs}
\usepackage{makecell}
\usepackage[numbers,sort&compress]{natbib}
\usepackage{graphicx}%
\usepackage{subfigure}
\usepackage{epsfig}
\usepackage{epstopdf}
\graphicspath{{./figures/}}%

\usepackage{abstract}
\usepackage{setspace}
\usepackage{section}
\usepackage{natbib}
\usepackage[sort&compress]{natbib}

\ifCLASSOPTIONcompsoc
\usepackage[caption=false,font=normalsize,labelfont=sf,textfont=sf]{subfig}
\else
\usepackage[caption=false,font=footnotesize]{subfig}
\fi

\begin{document}
%

\title{Ocean Reverberation Suppression via Matrix Completion \\ with Sensor Failure}
%
%
%

\author{
\normalsize
Li-ya~XU$^{1}$, Bin~LIAO$^{1}$, Hao~ZHANG$^{2}$,  Peng~XIAO$^{1}$,Jian-jun~HUANG$^{1}$\\
$^{1}$ College of Electronics and Information Engineering, Shenzhen University,  \\
Shenzhen 518060, China \\
$^{2}$ School of Marine Science and Technology,
Northwestern Polytechnical University,\\Xi’an 710072, China\\
Corresponding author:  Peng~XIAO, llsywan@hotmail.com
\date{}
}

\maketitle

\pagestyle{empty}  
\thispagestyle{empty} 
\normalsize
\renewcommand{\abstractnamefont}{\Large\bfseries}
\begin{abstract}
\normalsize

As one of the most important physical phenomena of underwater acoustics, ocean reverberation is a common and strong interference which  significantly degrades the performance of target bearing estimation. Meanwhile, sensor failure is inevitable in actual sonar deployment as the underwater scene is complicated. Therefore, it is a challenge for acoustic localization in the ocean reverberation with sensor failure. To address this issue, we propose an improved approach based on the principle of low rank characteristics of matrix in this paper. 
Firstly, we utilize Hankel structured matrix to counteract the problem of sensor failure. Then, the algorithm of matrix completion (MC) based on  $\ell_{1}$-norm and $\ell_{2}$-norm are {exploited}  to recover the true signal matrix from the corrupted received matrix.
Numerical results demonstrate that  compared with  other related methods,  the $\ell_{1}$-norm  provides better capability in the probability of recovery and shows the best robustness of bearing estimation with the narrow beam width and low sidelobe. {Moreover, the proposed approach verifies} the performances of reverberation suppression and {achieves} high resolution of localization. 
\end{abstract}


\IEEEpeerreviewmaketitle

\section{Introduction}
\begin{figure}[ht]
   \centering
  \includegraphics[width=0.45\textwidth]{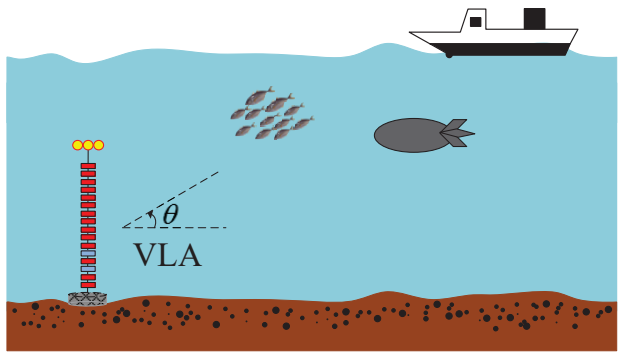} %
  \caption{DOA estimation with a VLA in ocean reverberation environment.} 
  \label{Figl} 
\end{figure}

There exist various irregular scatterers in the ocean,  such as marine lives, sediment particles, air bubbles, water masses, as well as roughness of sea surface and seafloor. The inhomogeneity generated by all these random scatterers forms a scattering field \citep{Ref1}.
Similar to the long, slow and quivering sounds, ocean reverberation  caused by scattering is one of the most important physical phenomena of underwater acoustics. According to the scattering sources, ocean reverberation can be divided into three categories, i.e., volume reverberation, sea surface reverberation, and seafloor reverberation \citep{Ref2}. Besides the ambient noise, ship noise, and so on, reverberation is known to be a common and strong interference in source localization. It is generated along with the transmit signal, and hence is closely related to the transmit signal itself. Additionally, it is also related to the propagation characteristics of sound field.
Usually, the target signal is overwhelmed by reverberation, which severely limits the effective working distance and degrades the performance of sonar system. Fig. \ref{Figl}  shows the the acoustic localization with a vertical line array (VLA) in ocean reverberation environment.
As reverberation is formed by the superposition of a large number of random scatterers at the  received sonar, it is a random process. In general, it is challenging to suppress reverberation  under ocean reverberation background. 

Many works have investigated reverberation suppression to improve the constant false alarm probability, especially for weak targets in the ocean.
When combining the reverberation  with  transmit signals,  the transmit waveform can be designed to reduce the influence of reverberation on the target, but it still needs to be analyzed by signal processing algorithms to extract the actual waveforms information \citep{Ref3,Ref4}. 
At the receiver, the reception directivity can be improved by increasing the azimuth direction aperture for reverberation suppression \citep{Ref5,Ref6} . According to the theory of normal mode, the acoustic signals can be composed of different modes. The reverberation usually has a high-order mode, and the filtering of normal mode can improve the signal-to-reverberation ratio (SRR), where a group or a low-order mode is chosen to enhance the target signal.
The disadvantage of this method is that the choice of the normal mode heavily depends on the ocean environment and source depth \citep{Ref7,Ref8}. Besides, the background of reverberation can be normalized by adaptive beamforming algorithm \citep{Ref5,Ref9,Ref10}. However, it requires efficient estimation of the correlation matrix, and the estimation error will increase because of the instability. 
In addition, the classic matched processing method can be performed to use the transmit signal as a reference for coherent compression. Nevertheless, this method only performs well in white noise background. 
Therefore, a generalized matched filtering method \citep{Ref11} is firstly modified to whiten the reverberation by using the parametric model filtering with a recursive procedure. Meanwhile, the copy of sound field and the optimum focusing filtering are mainly used to reduce the mismatch error. 
Furthermore, the concept of time-reversal in acoustics is also derived from the method of phase conjugate in the frequency domain in optics. \cite{Ref12} used the time-reversal process to focus the acoustic energy on the target position,  thereby effectively enhancing the target echo to improve the SRR. 
\cite{Ref13,Ref14} proposed a time-reversal processing based on reverberation groove method and carried out an experimental demonstration. However, the zero point of reverberation in the specified distance and the excitation of the time-reversal array are not set. 
Due to the sparse property in detection circumstances, 
compressed sensing (CS) is utilized for target estimation \citep{Ref15}. It is demonstrated that the results of estimation achieved by CS beamforming have a higher azimuth resolution and higher detection ability than the conventional beamforming methods \citep{Ref16, Ref17, Ref18}. 
The detection of sparse target can be implemented in fractional Fourier transform, which exhibits an improved localization performance \citep{Ref19, Ref20}. Moreover, based on the difference between the reverberation and target signal, the method of principal component analysis (PCA) is used to suppress the reverberation. It does not require prior statistical knowledge and deterministic models, and hence it is highly adaptable  \citep{Ref21, Ref22}.

DOA estimation is one of the main approaches for acoustic localization. Note that the array processing algorithms in the ocean are generally based on the assumption of absence of errors in the hydrophone. Undesirably, in practical applications, due to the effects of hydrophone technologies, deployment methods, device aging and ocean environment, corrupted hydrophones with sensor failure are inevitable.
Fig. \ref{Figl} shows two corrupted hydrophones (marked with blue) in the VLA. The corrupted hydrophones will destroy the geometric distribution structure of the array. For example, a corrupted element will make the uniform line array  lose the geometric symmetry, thereby reducing the array gain. 
The most direct method is to replace the corrupted hydrophones. However, it is difficult to implement when the sonar array has been placed in water because the acoustic data cannot be easily obtained in the experiment. 
Accordingly, reconstruction algorithms of the acoustic data can be used to reduce the influence of the corrupted sonar array and avoid the human and material resources incurred by replacing the hydrophones. Moreover, if the probability of corrupted array is taken into account in the early stage of data processing, the effect on array signal processing can be prevented. 

For the corrupted array, there are the weight re-optimization of the remaining normal array elements and the reconstruction of contaminated array signals. According to the desired array pattern, the weight of the remaining normal array elements can be re-optimized, so that the directional pattern formed by the effective elements is as close as possible to the original one. 
The sidelobe level caused by the corrupted array elements can be suppressed by conjugate gradient method \citep{Ref23}, quadratic constrained linear optimization \citep{Ref24}, maintaining fixed nulls \citep{Ref25}, adaptive weight reconstruction \citep{Ref26}, and orthogonal  methods \citep{Ref27}. 
To overcome the limitations on a long distance and huge computing quantity, these methods of weight re-optimization are not easy to achieve as all the weights must be readjusted. The basic idea of signal reconstruction is to use the received data of the normal array elements to reconstruct the data of the corrupted array elements by interpolation or optimization algorithms. For a uniform linear array, the signal received by adjacent array elements has only a fixed phase difference, so the output of corrupted array elements can shift the  phase  according to the normal array element output. 
This method is relatively simple for single source but complicated for multiple sources. Generally, adjacent \citep{Ref28} or all \citep{Ref29} normal array elements can be used to reconstruct the output signal. Besides bilinear programming \citep{Ref30}, cumulants \citep{Ref31}, inverse Fourier transform \citep{Ref32}, and global optimization algorithms (such as genetic algorithm \citep{Ref33}, particle swarm optimization algorithm \citep{Ref34} and neural network \citep{Ref35}) have been successfully applied to reconstruct the corrupted elements.

The remainder of this paper is organized as follows. Section II presents the problem formulation.  Section III shows the proposed solution based on $\ell_{p}$-norm.  Section IV introduces the application of our algorithm under reverberation and  the superiority over other related algorithms. Finally,  Section V draws conclusions.

\section{Problem Formulation}

Assume there are  $r$  underwater targets from unknown distinct direction of $\{\theta_1,\cdots,\theta_r\}$ impinging on the VLA which has $N$ sensors with array spacing of half-wave. Moreover, we use  $\boldsymbol{A} (\theta)\in \mathbb{C}^{N\times r}$ to denote the steering matrix of all the steering vectors. It is also assumed that the DOAs of the source are not time-varying and the number of snapshots is $M$. Then,  we have
\begin{equation}
\boldsymbol{X}=(\boldsymbol{A}(\theta)\boldsymbol{S}+\boldsymbol{R})^{T}.
\label{equation1}
\end{equation}
In \eqref{equation1}, $\boldsymbol{X}=\left[ \boldsymbol{x}_{1}, \boldsymbol{x}_{2}, \ldots, \boldsymbol{x}_{N}\right] \in \mathbb{C}^{M\times N}$ is the array data matrix, and the $n$th element ${\boldsymbol x}_n$ of $\boldsymbol{X}$ denotes  the samples of  sensor  $n$. Moreover,  $\boldsymbol{S}=\left[ \boldsymbol{s}_{1}, \boldsymbol{s}_{2}, \ldots, \boldsymbol{s}_{M}\right] \in \mathbb{C}^{r\times M} $ is the signal matrix, where  $\boldsymbol{s}_{m}$ represents the signal vector at time instant $m$. $\boldsymbol{R}\in \mathbb{C}^{N\times M} $ is the measured reverberation matrix. In practical systems, we usually have $ r<N<M $.  Therefore,  $\boldsymbol{M}=\boldsymbol{A}(\theta)\boldsymbol{S}\in \mathbb{C}^{N\times M }  $  is a low-rank matrix. 

\begin{figure}[ht]
\centering
\subfigure[  ]{               
\includegraphics[width=0.4\linewidth]{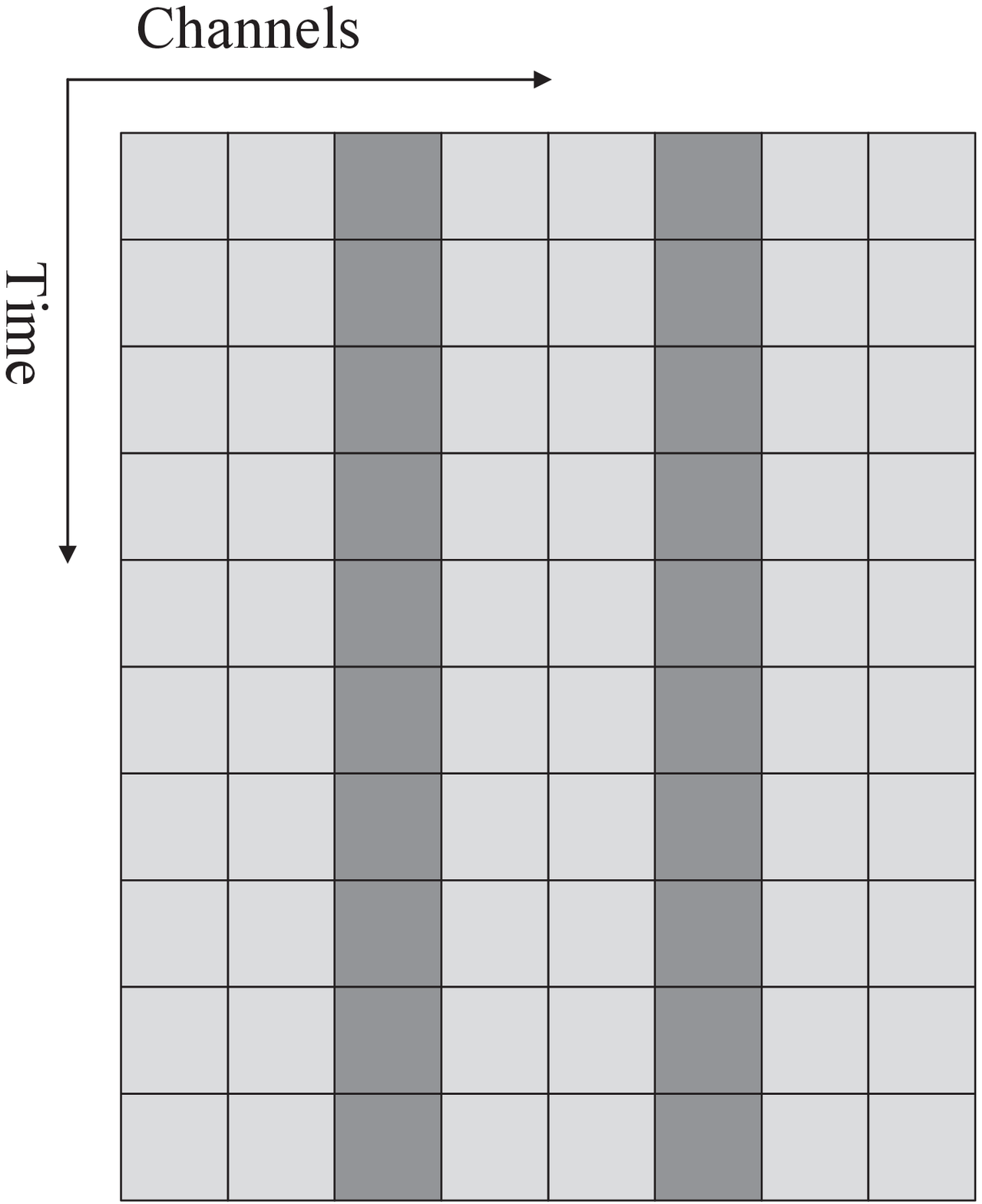}}
\hspace{2mm}
\subfigure[  ]{
\includegraphics[width=0.4\linewidth]{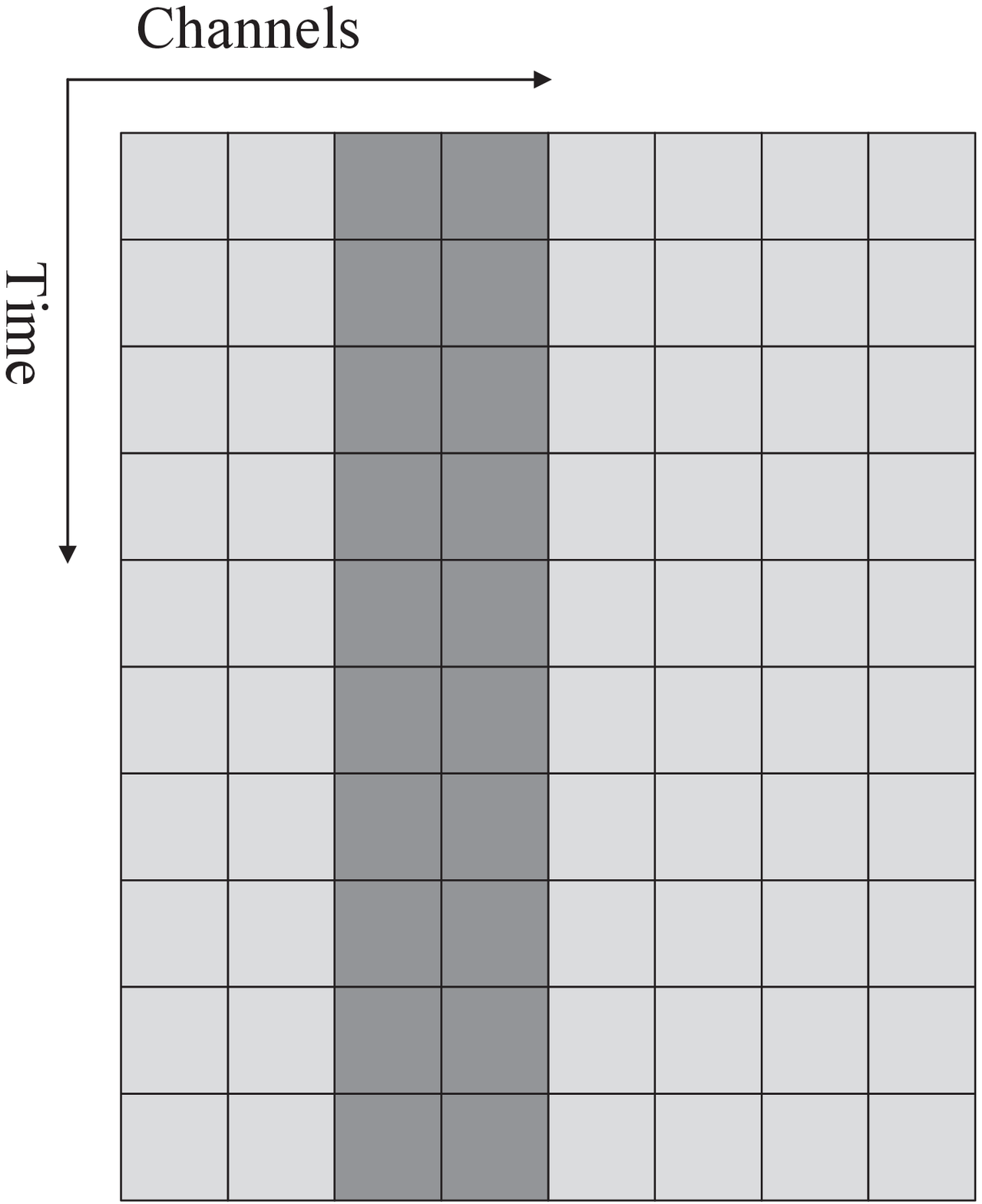}}
\centering\caption{The model of sensor failure in sonar array.}
\label{Fig:2}
\end{figure}

In order to model sensor failure in sonar array, the following two cases illustrated in Fig. \ref{Fig:2} will be investigated:
(a) Some hydrophones are failed during the entire sampling period, and the missing positions of channel are randomly selected.
(b) Some hydrophones are failed during the entire sampling period, and the missing positions of channel are randomly and continuously selected.
When the array is corrupted with sensor failure, the incomplete measurement matrix can be written as
\begin{equation}
\boldsymbol{X}=\boldsymbol{B} \odot (\boldsymbol{M}+\boldsymbol{R})^{T}.
\label{equation2}
\end{equation}
In \eqref{equation2},  $\boldsymbol{B}  $ is a binary matrix composed of 0 and 1 at the positions associated with normal and corrupted data, respectively. $\odot$  denotes element-wise product. In this paper, we assume $\boldsymbol{B}$  has been judged by the received acoustic data. Moreover, we define the operator $\mathcal{P}_{\Omega} $ as  $ [\mathcal{P}_{\Omega}(\boldsymbol{X})]_{i,j} = X_{i,j} $  if  $(i,j) \in {\Omega} $  and 0 otherwise, where  $\Omega \in \{ 1,...,M \} \times \{ 1,...,N \} $ contains the indices of observed entries.

With the improvement of sonar resolution, it leads a fact that the number of effective ocean scatterers in each spatial processing unit is greatly reduced, which no longer meets the central limit theorem in statistics.
The  envelope of reverberation $\boldsymbol{R}_{e}(x)$  of high-resolution sonar has serious smearing. Therefore, it is with skewness and no longer obeys the Gaussian distribution. 
So, we can regard the ocean reveberation as outliers in the receiverd matrix.
The statistical properties have been investigated by some experimental data, and in this paper we choose the $K$ distribution to describe the reverberation \citep{Ref44,Ref45,Ref46} which can be determined by different degrees of freedom $n$. Moreover, the phase  $\boldsymbol{R}_{p}(\varphi)$ can be described by the uniform distribution:
\begin{equation}
\boldsymbol{R}_{e}(x)= \left\{
\begin{array}{rl} {\frac{1}{2^{n / 2} \Gamma(n / 2)} x^{\frac{n}{2}-1} \mathrm{e}^{-x / 2},}& { \textrm{if}~~~~~x>0} \\
{0,}&  \textrm{else}
\end{array}
\right. ,
\end{equation}
\begin{equation}
\boldsymbol{R}_{p}(\varphi)=\left\{
\begin{array}{rl}{\frac{1}{2\pi},} & { \textrm{if}~~~~\varphi \in(0, 2\pi)}\\
{0,}& \textrm{else}
\end{array}
\right. .
\end{equation}
Accordingly, the reverberation signal $\boldsymbol{R}$  in \eqref{equation1} can be obtained by
\begin{equation}
\boldsymbol{R}=\boldsymbol{R}_{e}(x)e^{j\boldsymbol{R}_{p}(\varphi) }.
\end{equation}

In the absence and corruption of data, it is known that PCA is a common approach to data analysis and dimensionality reduction, e.g., information extraction such as DOA estimation. However, if the entries in the observation matrix are severely damaged, the error between the matrix obtained by PCA and the original matrix will be very large. 
This is mainly because PCA is sensitive to outliers. The accurate calculation of principal components in the presence of outliers is called  robust principal component analysis (RPCA) \citep{Ref36,Ref37}. The MC \citep{Ref38} theory proposed by Cand$\rm\grave{e}$s \textit{et al.} is an important method for data analysis and processing after the CS theory. 
CS and MC are the main components of principal component pursuit (PCP) \citep{Ref39,Ref40}, which can be used to solve RPCA. When the data matrix is severely damaged or corrupted by noise, MC is capable of the near-perfect recovery of matrices if the original matrices have low rank characteristics.
Motivated by the advantages of  MC techniques, in this paper, we propose to utilize the low-rank matrix factorization to recover the acoustic data caused by sonar sensor failure in ocean reverberation. Note that the conventional techniques for MC depend on the assumption of Gaussian interference; however, in the ocean environment, the background of noise is usually non-Gaussian. 
Zeng and So \citep{Ref41} have proved that $\ell_{p}$-norm with $0<{p}<2$ shows a better performance than the Frobenius norm in the presence of outliers. Thus,  with the reverberation signal in simulation,   $\ell_{1}$-norm and $\ell_{2}$-norm are utilized to reply to reverberation suppression, and Hankel structured matrix is employed to deal with sensor failure.

From  the MC theory we know that when the entries of a certain  row or column of a matrix are completely missing, generally, the missing entries cannot be directly recovered based on other entries. One solution is converting the matrix into a Hankel structured matrix, and then applying the MC theory to recover the missing entries. Let  $\mathcal{H}_{n_1} :\boldsymbol{X}\in \mathbb{C}^{M\times N} \to \boldsymbol{Z}\in \mathbb{C}^{M n_{1} \times {n_{2}}} $ denote a linear operator, mapping the data matrix into the corresponding Hankel matrix \citep{Ref42}:
\begin{equation}
 \mathcal{H}_{n_{1}}\left(\boldsymbol{X}\right)=\left[\begin{array}{cccc}{\boldsymbol{x}_{1}} & {\boldsymbol{x}_{2}} & {\cdots} & {\boldsymbol{x}_{n_{2}}} \\ {\boldsymbol{x}_{2}} & {\boldsymbol{x}_{3}} & {\cdots} & {\boldsymbol{x}_{n_{2}+1}} \\ {\vdots} & {\vdots} & {\ddots} & {\vdots} \\ {\boldsymbol{x}_{n_{1}}} & {\boldsymbol{x}_{n_{1}+1}} & {\cdots} & {\boldsymbol{x}_{N}}\end{array}\right],
\end{equation}
where $ n_{1}+n_{2}=N+1$, and  we usually take $n_{1}=N/2 $. It is not difficult to find that even if some columns of the original data matrix are missing, in the new reshaped Hankel matrix, the coordinates of the missing entries have been all shuffled, so the situation of all missing entries of some columns will no longer occur. In our paper, for simplicity, we use $\mathcal{H}(\boldsymbol{X})$  instead of $ \mathcal{H}_{n_{1}}\left(\boldsymbol{X}\right)$. For example, we assume that an original matrix is
\begin{equation}
\boldsymbol{X}=\begin{bmatrix} 1& 4 & 7& 10   \\ 2 & 5 & 8& 11 \\ 3 & 6 & 9& 12 \end{bmatrix}.
\label{equation7}
\end{equation}
If  column 2 of $\boldsymbol{X}$  corrupts, the unobserved entries over  $\Omega $ is  $\{ (1,2),(2,2),(3,2) \}$. Then,  we have
\begin{equation}
\mathcal{P}_{\Omega}(\boldsymbol{X})=\begin{bmatrix} 1&\times& 7& 10   \\ 2 & \times & 8& 11 \\ 3 & \times & 9& 12 \end{bmatrix},
\end{equation}
whose Hankel structured matrix is
\begin{equation}
\mathcal{H}(\mathcal{P}_{\Omega}(\boldsymbol{X})) =\begin{bmatrix} 1&\times& 7\\ 2 & \times & 8\\ 3 & \times & 9 \\ \times & 7 & 10 \\ \times & 8 & 11 \\ \times & 9 & 12   \end{bmatrix}.
\label{equation9}
\end{equation}
After the Hankel transformation, $\Omega$ $\to$  $\Omega^{'}$, 
the unobserved entries over $\Omega^{'} $ are  $\{ (1,2),$ $(2,2),$ $(3,2),$ $(4,1),$ $(5,1),$ $(6,1) \}$.
Accordingly, the inverse Hankel transformation of $ \mathcal{H}$,  i.e.,  $\mathcal{H}^{\dagger} $,   $\mathcal{H}^{\dagger}(\boldsymbol{Z})\in \mathbb{C}^{M\times N} $  \citep{Ref42} is given by 
\begin{equation}
\left(\mathcal{H}^{\dagger}(\boldsymbol{Z})\right)_{i, j}=\frac{1}{w_{j}} \sum\limits_{k_{1}+k_{2}=j+1} Z_{(k_{1}-1) n_{c}+i, k_{2}},
\label{equation10}
\end{equation}
where $w_j$  is the number of elements in the $j$th  anti-diagonal. For example,  the $3$rd anti-diagonal contains three elements over $\{ (3,1),$ $(2,2),$ $(1,3)\}$, so  $w_3 =3$;  and the $N$th anti-diagonal contains only one elements, so $w_N =1$.

The task of MC in this paper aims at recovering $\boldsymbol{M} $  from $\mathcal{P}_{\Omega}(\boldsymbol{X}) $, and the corresponding optimization problem is  formulated as
\begin{equation}
\begin{split}
&{\underset{\boldsymbol{M},\boldsymbol{R}} {\operatorname{min}}} \quad
{\parallel \mathcal{H}(\mathcal{P}_{\Omega}(\boldsymbol{X}))- \mathcal{H}(\mathcal{P}_{\Omega}(\boldsymbol{M}^{T}))\parallel_{p}^{p}}   \\
&\ {\text{s.t.} \quad\operatorname{rank}(\mathcal{H} (\boldsymbol{M}^{T})) } \leq r \\
\end{split},
\label{equation11}
\end{equation}
where $\Vert \cdot \Vert_{p} $ is the $ l_{p} $  norm of a matrix, i.e.,
\begin{equation}
\Vert \boldsymbol{E}_{\Omega}\Vert_{p}=\left(\sum\limits_{(i, j) \in \Omega} \Vert \boldsymbol{E}_{i,j} \Vert ^{p}\right)^{1/p}.
\label{def_LP_norm}
\end{equation}
In \eqref{def_LP_norm},  $\boldsymbol{E}_{\Omega} = \mathcal{H}(\mathcal{P}_{\Omega}(\boldsymbol{X}))- \mathcal{H}(\mathcal{P}_{\Omega}(\boldsymbol{M}^{T})) $ denotes the error matrix. It is worth mentioning that we  consider two situations $p=1 $  and $p=2 $ in this paper.
As the solution procedure is executed based on the Hankel structure, problem \eqref{equation11} is equivalent to
\begin{equation}
\begin{split}
&{\underset{\boldsymbol{M},\boldsymbol{R}} {\operatorname{min}}} \quad
{\parallel (\mathcal{H}(\boldsymbol{X}))_{\Omega^{'}}-  (\mathcal{H}(\boldsymbol{M}^{T}))_{\Omega^{'}}\parallel_{p}^{p}}   \\
&\ {\text{s.t.} \quad \operatorname{rank}(\mathcal{H} (\boldsymbol{M}^{T})) } \leq r \\
\end{split}.
\label{equivalent_opti_problem}
\end{equation}
The minimization with the observed entries is made a transitions from $\Omega $  to  $\Omega^{'} $. 

\begin{algorithm}[t]
\caption{: Hankel Structured MC Based on $\ell_{p}$-norm}
\label{alg:A}
\begin{algorithmic}[1]
\REQUIRE $\boldsymbol{X}_{\Omega}$,  $\Omega$, $\boldsymbol{B}$, $r$ and the iteration times $K$.\\
 \STATE  $\mathbf { Preprocess: }{ \boldsymbol{X} \to \mathcal{H}(\boldsymbol{X})}$.\\
 \STATE$\qquad \qquad \qquad \Omega \in \mathbb{C}^{M \times N } \to  \Omega^{'} = \mathcal{H}(\boldsymbol{B}) \in \mathbb{C}^{M n_1 \times n_2}$.
 \STATE${\mathbf { Initialize:}\text { Randomly initialize } \boldsymbol{U}^{0} \in \mathbb{R}^{Mn_{1} \times r}} $.
 \STATE${\text {Determine }\left\{\mathcal{I}_{jj}\right\}_{jj=1}^{n_{2}} \text { and }\left\{\mathcal{J}_{ii}\right\}_{ii=1}^{Mn_{1}} \text { according to } \Omega^{'}} $.\\
    \FOR{$k=0,1,\cdots,K$ }
      \FOR{$jj=1,2, \cdots, n_{2} $ }
        \STATE $(\boldsymbol{v}_{jj})^{k+1} \leftarrow $ $ \arg  \underset{\boldsymbol{v_{jj}}} {\operatorname{min}}$  $\parallel \boldsymbol{b}_{\mathcal{I}_{jj}} -\boldsymbol{U}_{\mathcal{I}_{jj}}^{k} \boldsymbol{v}_{jj} \parallel _{p}^{p}$.
        \ENDFOR
        \FOR{$ii=1,2, \cdots, Mn_{1} $ }
        \STATE $ (\boldsymbol{u}_{ii}^{T})^{k+1} \leftarrow $ $\arg  \underset{\boldsymbol{{u}_{ii}^{T}}} {\operatorname{min}} $ $\parallel \boldsymbol{b}_{\mathcal{J}_{ii}}^{T}     -\boldsymbol{u}_{ii}^{T} \boldsymbol{V}_{\mathcal{J}_{ii}}^{k+1}\parallel_{p}^{p}$.
       \ENDFOR
    \ENDFOR
\ENSURE $\boldsymbol{M}=\mathcal{H}^{\dagger} ({\boldsymbol{U}^{k+1} \boldsymbol{V}^{k+1}}) $.
\end{algorithmic}
\end{algorithm}

\section{Hankel structured MC based $\ell_{p}$-norm }
Low rank matrix factorization \citep{Ref41}, which is considered as the low rank property of matrix, has been used to void full singular value decomposition (SVD). Therefore, optimization problem \eqref{equivalent_opti_problem}  can be transformed into 
\begin{equation}
{\underset{\boldsymbol{U},\boldsymbol{V}} {\operatorname{min}}} \quad f_{p}({\boldsymbol{U},\boldsymbol{V}} ) :=\parallel ( \mathcal{H}(\boldsymbol{X}))_{\Omega^{'}}  - (\boldsymbol{U}\boldsymbol{V})_{\Omega^{'}}  \parallel_{p}^{p},
\end{equation}
where $ \boldsymbol{U}\in \mathbb{C}^{Mn_{1}\times r}$  and $ \boldsymbol{V}\in\mathbb{C}^{r\times n_{2}}  $. Because  $r\ll{Mn_{1}}{n_{2}} $, the complexity of MC  will be   greatly reduced. Furthermore, the smaller the rank $r$, the higher the accuracy of MC. The error matrix after Hankel structured processing can be rewritten as $\boldsymbol{E}_{\Omega^{'}} = ( \mathcal{H}(\boldsymbol{X}))_{\Omega^{'}} -(\boldsymbol{U}\boldsymbol{V})_{\Omega^{'}} $. It is worth noting that the inverse Hankel transformation in \eqref{equation10} needs to be operated with $\widehat{\boldsymbol{M}}=(\mathcal{H}^{\dagger}(\boldsymbol{\boldsymbol{U}\boldsymbol{V}}))^{T} $ when $\boldsymbol{U} $  and $\boldsymbol{V} $   have been determined.

The strategy of alternating minimization with a relaxation as a bi-convex problem is adopted to solve $\boldsymbol{U} $  and $\boldsymbol{V} $ respectively according to:
\begin{equation}
{\boldsymbol{V}^{k+1}} =\arg \underset{\boldsymbol{V}} {\operatorname{min}} {\parallel ( \mathcal{H}(\boldsymbol{X}))_{\Omega^{'}}-(\boldsymbol{U}^ {k}\boldsymbol{V})_{\Omega^{'}}\parallel_{p}^{p}} ,
\end{equation}
\begin{equation}
{\boldsymbol{U}^{k+1}} = \arg \underset{\boldsymbol{U}} {\operatorname{min}} {\parallel ( \mathcal{H}(\boldsymbol{X}))_{\Omega^{'}}-(\boldsymbol{U}\boldsymbol{V}^{k+1})_{\Omega^{'}}\parallel_{p}^{p}} .
\end{equation}

In the first step, we solve $\boldsymbol{V} $  by fixing $\boldsymbol{U} $, i.e.,
\begin{equation}
{\underset{\boldsymbol{V}} {\operatorname{min}}} \ f_{p}({\boldsymbol{V}} ) :={\parallel ( \mathcal{H}(\boldsymbol{X}))_{\Omega^{'}}  - (\boldsymbol{U}\boldsymbol{V})_{\Omega^{'}}  \parallel_{p}^{p}},
\label{equation17}
\end{equation}
where $\left( \cdot \right)^{k} $  is omitted for simplicity. Let $\boldsymbol{u}_{ii}^{T} \in \mathbb{C}^{ r}$  and  $\boldsymbol{v}_{jj}  \in \mathbb{C}^{r }$ denote the $ii$th row of $\boldsymbol{U} $  and the $jj$th column of $\boldsymbol{V} $, respectively, where $ii=1,...,Mn_1   $ and $jj=1,...,n_{2}$.
Then, problem \eqref{equation17} can be expressed as follow:
\begin{equation}
{\underset{\boldsymbol{V}} {\operatorname{min}}} \ f_{p}({\boldsymbol{V}} ) :=\sum\limits_{(ii,jj)\in\Omega{'}}{\left| ( \mathcal{H}(\boldsymbol{X}))_{ii,jj}  - \boldsymbol{u}_{ii}^{T}\boldsymbol{v}_{jj}  \right|^{p}}.
\label{equation18}
\end{equation}
According to the number of columns in $\boldsymbol{V} $, problem  \eqref{equation18} can be decomposed into $n_{2} $  independent subproblems:
\begin{equation}
{\underset{\boldsymbol{v}_{jj}} {\operatorname{min}}} \ f_{p}({\boldsymbol{v}_{jj}} ) :=\sum\limits_{ii \in \mathcal{I}_{jj}}{\left| ( \mathcal{H}(\boldsymbol{X}))_{ii,jj}  - \boldsymbol{u}_{ii}^{T}\boldsymbol{v}_{jj}  \right|^{p}}.
\label{equation19}
\end{equation}
Here we use $\mathcal{I}_{jj} =\left\{ jj_1,...,jj_{\mathcal{I}_{jj}} \right\}$ and $\vert \mathcal{I}_{jj} \vert$ to denote  the  $jj $th column of $\Omega^{'} $ and its cardinality, respectively, where  $\vert \mathcal{I}_{jj} \vert>r$. From  the  Hankel structured matrix given in \eqref{equation9}, we know that  only the  $(1,1)$,  $(2,1)$ and $(3,1)$ can be  observed when $jj=1 $, so we have  $\mathcal{I}_{1}=\{1,2,3\}$ and  $\mathcal{I}_{2}=\{4,5,6\} $.

Furthermore, we respectively define a matrix $\boldsymbol{U}_{\mathcal{I}_{jj}} \in \mathbb{C}^{\mathcal{I}_{jj}\times  r}$ that contains the observed row indexed by  $\mathcal{I}_{jj} $:
\begin{equation}
\boldsymbol{U}_{\mathcal{I}_{jj}}=\left[\begin{array}{c}{\boldsymbol{u}_{jj_{1}}^{T}} \\
{\vdots} \\
 {\boldsymbol{u}^{T}_{jj_{\vert{\mathcal{I}_{jj}\vert}}}}\end{array}\right],
\end{equation}
and a vector $\boldsymbol{b}_{\mathcal{I}_{jj}} =[\boldsymbol{X}_{jj_1,jj},...,\boldsymbol{X}_{jj_{\mathcal{I}_{jj}},jj} ]^{T} \in \mathbb{C}^{\mathcal{I}_{jj} }$. Then, problem \eqref{equation19} is equivalent to
\begin{equation}
{\underset{\boldsymbol{v}_{jj}} {\operatorname{min}}} \ f_{p}({\boldsymbol{v}_{jj}} ) :={\Arrowvert \boldsymbol{b}_{\mathcal{I}_{jj}}  - \boldsymbol{U}_{\mathcal{I}_{jj}}\boldsymbol{v}_{jj}  \Arrowvert_{p}^{p}}.
\label{equation21}
\end{equation}
When $p=2$, problem \eqref{equation21} becomes a least squares problem and can be solved by ${\boldsymbol{v}_{jj}}=\boldsymbol{U}^{\dagger}_{\mathcal{I}_{jj}} \boldsymbol{b}_{\mathcal{I}_{jj}}$, and  the  implementary complexity of the $jj$th iteration  is $ \mathcal{O} \left(\vert \mathcal{I}_{jj} \vert r^{2} \right)$. When $p=1$, problem \eqref{equation21} can be solved by the iteratively reweighted least squares (IRLS) algorithm \cite{Ref43}. We use $N_{\mathrm{IRLS}}$ to definite the number of iteration of IRLS, and  the complexity of $jj$-iteration here is $ \mathcal{O} \left(\vert \mathcal{I}_{jj} \vert r^{2} N_{\mathrm{IRLS}}\right)$; moreover, for the $n_{2}$ independent subproblems, we have  $\sum_{jj=1}^{n_{2}}\left|\mathcal{I}_{jj}\right|=|\Omega^{'}|$, so the overall complexity is $ \mathcal{O} \left(\vert\Omega^{'}\vert r^{2} N_{\mathrm{IRLS}}\right)$.

As the acoustic data is complex-valued, the real-valued transforms need to be preprocessed in the above algorithm. 
Then, the real-valued vector of  complex-valued $\boldsymbol{b}_{\mathcal{I}_{jj}} \in \mathbb{C}^{\mathcal{I}_{jj} }$  is
\begin{equation}
\boldsymbol{\boldsymbol{b}_{\mathcal{I}_{jj}}}_{_r}=\left[\begin{array}{c}{\Re\{\boldsymbol{\boldsymbol{b}_{\mathcal{I}_{jj}}}\}} \\ {\Im\{\boldsymbol{\boldsymbol{b}_{\mathcal{I}_{jj}}}\}}\end{array}\right] \in \mathbb{R}^{2 {\mathcal{I}_{jj} }}.
\end{equation}
Besides, the real-valued matrix of complex-valued $\boldsymbol{\boldsymbol{U}_{\mathcal{I}_{jj}}} \in \mathbb{C}^{\mathcal{I}_{jj} \times r}$ is
\begin{equation}
\boldsymbol{\boldsymbol{U}_{\mathcal{I}_{jj}}}_{_r}=\left[\begin{array}{cc}{\Re\{\boldsymbol{\boldsymbol{U}_{\mathcal{I}_{jj}}}\}} & {-\Im\{\boldsymbol{\boldsymbol{U}_{\mathcal{I}_{jj}}}\}} \\ {\Im\{\boldsymbol{\boldsymbol{U}_{\mathcal{I}_{jj}}}\}} & {\quad\Re\{\boldsymbol{\boldsymbol{U}_{\mathcal{I}_{jj}}}\}}\end{array}\right] \in \mathbb{R}^{(2 \mathcal{I}_{jj}) \times(2 r)}.
\end{equation}

Similarly, the  $ii$th row of $\boldsymbol{U} $  can be also solved trough  $M n_{1} $   independent subproblems:
\begin{equation}
{\underset{\boldsymbol{u}_{ii}^{T}} {\operatorname{min}}} \ f_{p}({\boldsymbol{u}_{ii}^{T}} ) :={\Arrowvert \boldsymbol{b}_{\mathcal{J}_{ii}}^{T}  - \boldsymbol{u}_{ii}^{T}\boldsymbol{V}_{\mathcal{J}_{jj}}  \Arrowvert_{p}^{p}},
\end{equation}
where $\boldsymbol{V}_{\mathcal{J}_{ii}}\in\mathbb{C} ^{r \times {\mathcal{J}_{ii}}}$  and  $\boldsymbol{b}_{\mathcal{J}_{ii}}^{T} =[\boldsymbol{X}_{ii,ii_1},...,\boldsymbol{X}_{ii,ii_{\mathcal{J}_{ii}}} ]^{T} \in \mathbb{C}^{\mathcal{J}_{ii} }$, and $\mathcal{J}_{ii} =\left\{ ii_1,...,ii_{\mathcal{J}_{ii}} \right\} \in \{1,...,n_2\}$  denotes the column for the $ii $th row in $\Omega^{'}$. In  \eqref{equation9}, when $ii=1$, we have  $\mathcal{J}_{1}=\mathcal{J}_{2}=\mathcal{J}_{3}={\{1,3\}} $  and $\mathcal{J}_{4}=\mathcal{J}_{5}=\mathcal{J}_{6}={\{2,3\}} $.

Above all, the proposed algorithm consists of two stages. The first stage preprocesses $\boldsymbol{X} $  into $ \mathcal{H}(\boldsymbol{X}) $, and then converts $\Omega$   into $\Omega^{'}$. The second stage fulfills  the MC with  $\ell_{p}$-norm in $\Omega^{'}$.
The overall complexity of the proposed algorithm is $ \mathcal{O} \left(\vert\Omega^{'}\vert r^{2} N_{\mathrm{IRLS}} K \right)$, where $K$ is  the  iteration times.

\section{Applications in reverberation suppression}
In this paper, we use the SRR defined as follow to compare the level of the target signal  with the reverberation:
\begin{equation}
\mathbf{SSR}=10 \log _{10}\left(\frac{ \sigma_{s}^{2}}{\sigma_{n}^{2}}\right) \mathrm{dB},
\end{equation}
where $\sigma_{s}^{2}  $ and $\sigma_{n}^{2} $ are the variances of signal and reverberation, respectively.

Moreover, the normalized root mean square error (RMSE) defined as follows is  exploited to measure the performance based on 100 independent trials: 
\begin{equation}
\mathbf{RMSE}(\widehat{\boldsymbol{M}})=\sqrt{\mathrm{E}\left\{\frac{\|\widehat{\boldsymbol{M}}-\boldsymbol{M}\|_{F}^{2}}{\|\boldsymbol{M}\|_{F}^{2}}\right\}},
\end{equation}
where $\widehat{\boldsymbol {M}}$  is the result obtained by equations (12) and (13). A typical experimental setting is  $N=20 $,  $M=100 $, and  $r=2 $. We compare the proposed algorithm respectively based on  $\ell_{1}$-norm and $\ell_{2}$-norm with OptSpace \citep{Ref47} and structured alternating projection (SAP) \citep{Ref42} to recover $\boldsymbol{M} $  from $\mathcal{P}_{\Omega}(\boldsymbol{X})$.

\begin{figure}[h]
   \centering
  \includegraphics[width=0.45\textwidth]{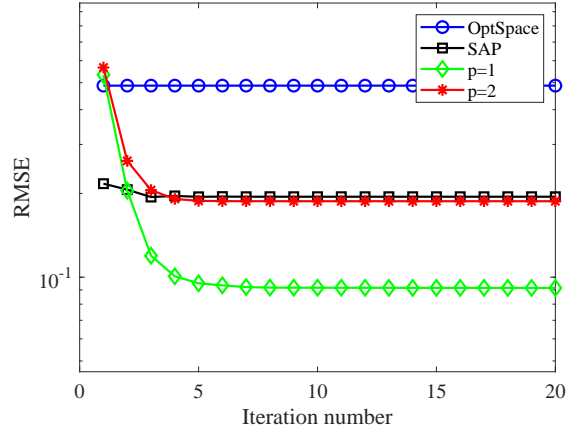} 
  \caption{The normalized RMSE versus iteration number.} 
  \label{Fig:3} 
\end{figure}

Fig. \ref{Fig:3} plots the normalized RMSE versus the iteration times. We set $\mathrm{SSR}=10 $, and the sensor failure percentage is $30\%$, and  the degrees of freedom of K distribution is  $n=0.1 $. We can observe that the OptSpace algorithm exhibits a high error without falling. Moreover, the SAP and  $\ell_{2}$-norm begin to converge and they have almost same the RMSEs.  Besides, they also provide  robust performances  in the ocean reverberation. However,  their performances are still worse than that of of the   $\ell_{1}$-norm.  Finally, it can be observed  that the algorithms of SAP,  $\ell_{1}$- and  $\ell_{2}$-norms converge within ten iterations.

\begin{figure}[h]
\centering
\subfigure[  ]{     
\includegraphics[width=0.45\linewidth]{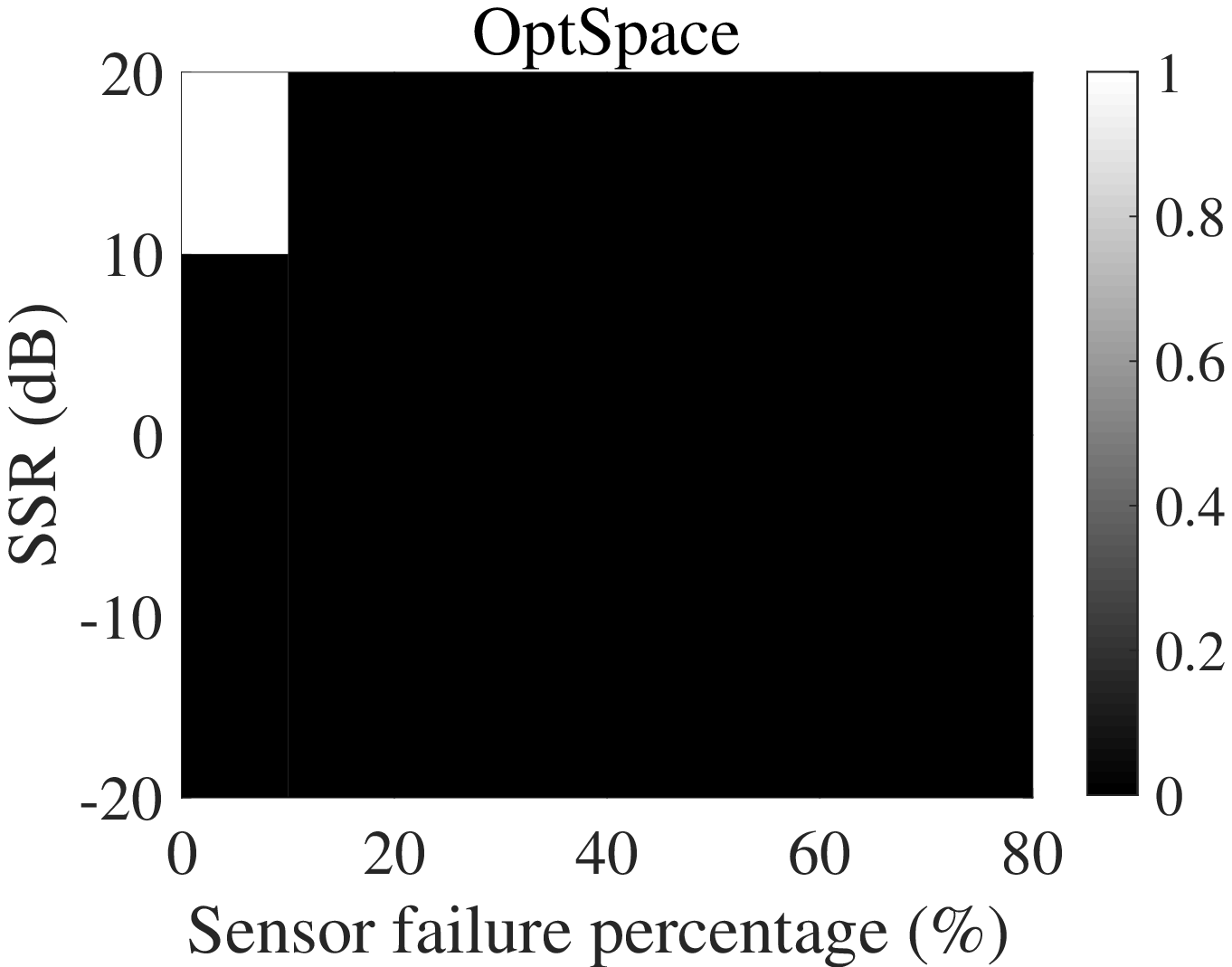}}
\hspace{0mm}
\subfigure[  ]{
\includegraphics[width=0.45\linewidth]{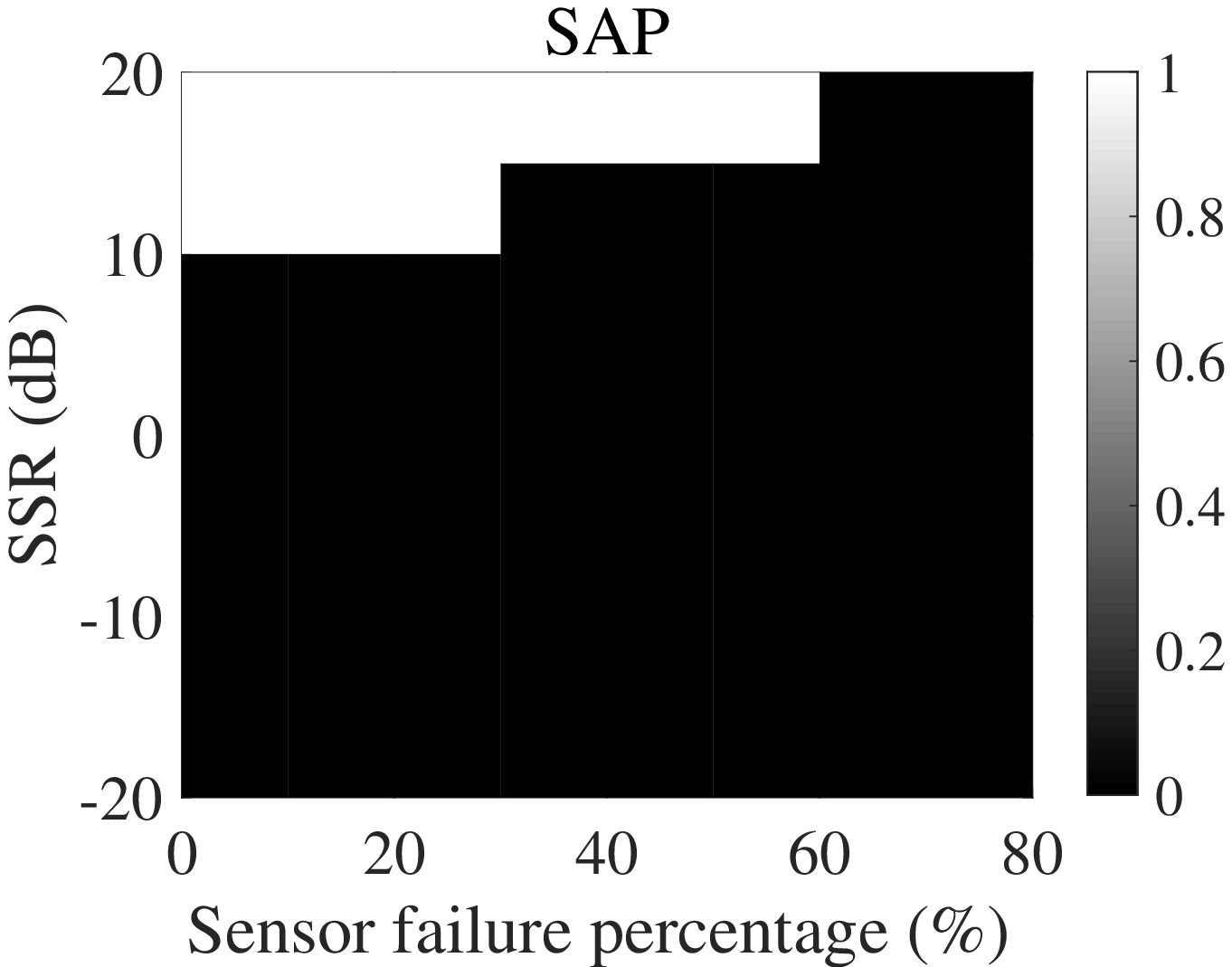}}
\hspace{0mm}
\subfigure[  ]{
\includegraphics[width=0.45\linewidth]{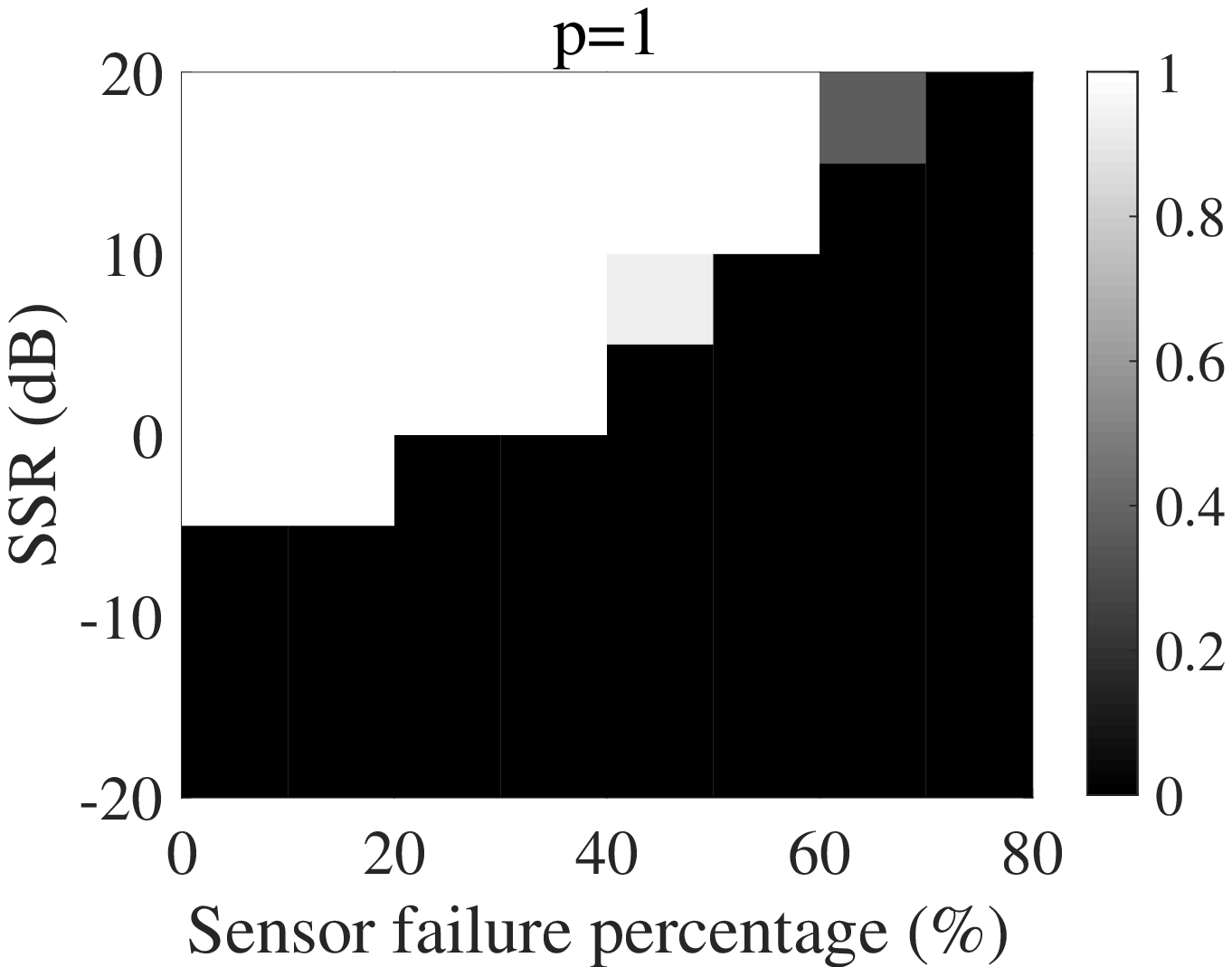}}
\hspace{0mm}
\subfigure[  ]{
\includegraphics[width=0.45\linewidth]{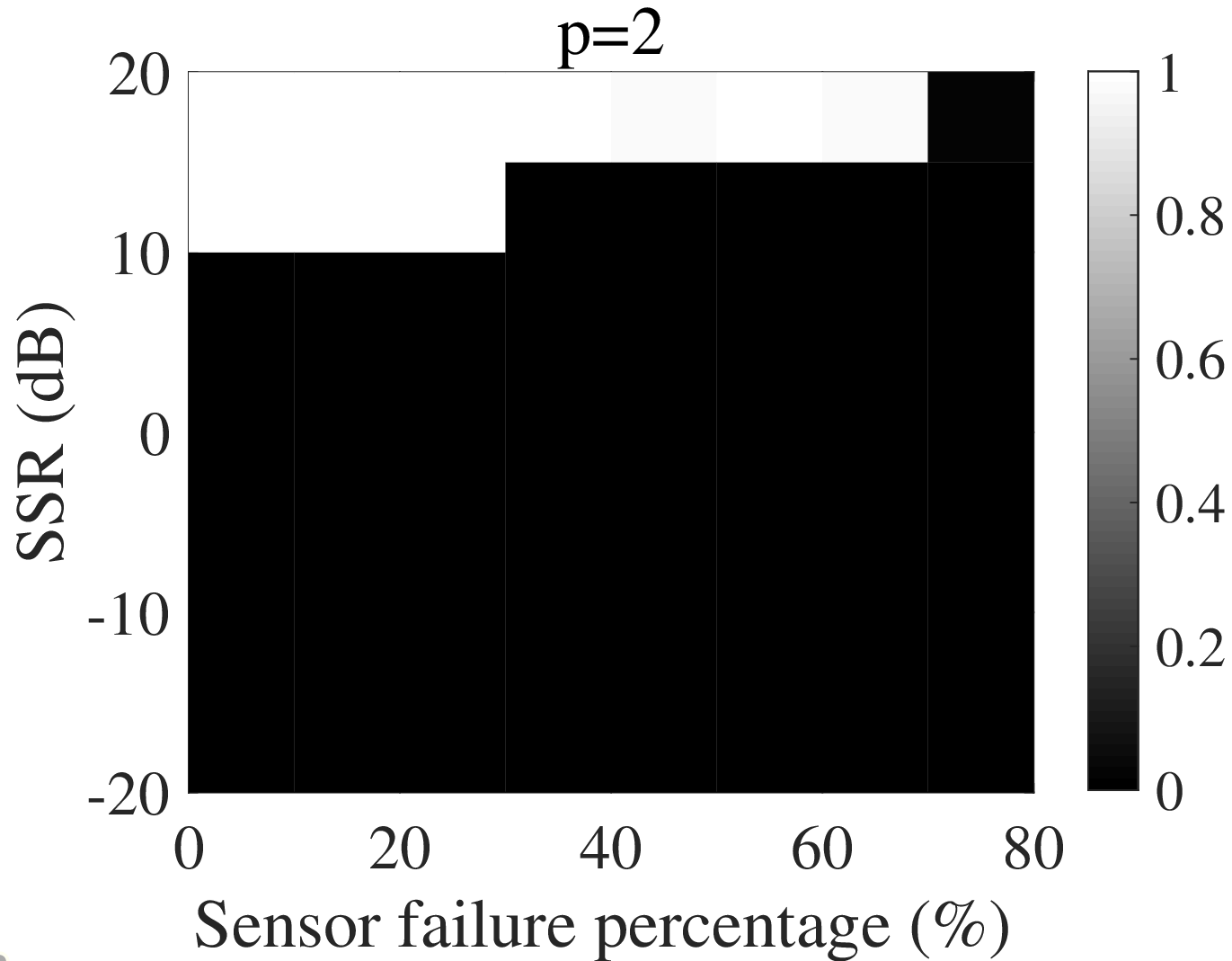}}
\hspace{0mm}
\centering\caption{The probability of recovery versus SSR and senor failure percentage.}
\label{Fig:4}
\end{figure}

Fig. \ref{Fig:4} illustrates the SSR with respect to the  probability of recovery. For each realization, 100 independent trials were evaluated. Firstly, if the normalized RMSE is smaller than 0.15, we think that the trial is successful and observed as white region, and dark region otherwise. Moreover, we can observe that the algorithm based on $\ell_{1}$-norm has the best performance than others in the ocean reverberation. 
Besides, the algorithm of OptSpace is only valid when the $\mathrm{SSR}$ is above 10 dB and the sensor failure percentage is smaller than $10\%$.
It is also worth noting that the SAP and  algorithm based on $\ell_{2}$-norm have similar  performances.  Therefore, the  MC algorithm based on  $\ell_{1}$-norm is suitable for the reverberation suppression.

In this paper,  the performance of DOA estimation is measured in terms of the RMSE:
\begin{equation}
\mathbf{RMSE}(\widehat{\boldsymbol{\theta}})=\sqrt{\frac{1}{100r} \sum\limits_{p=1}^{r}   \sum\limits_{q=1}^{100}    {(\widehat{\boldsymbol{\theta}}_{q,p}-\boldsymbol{\theta}_{q})^{2}}} ,
\end{equation}
where $\widehat{\boldsymbol {\theta}}$  is the estimated DOA.

\begin{figure}[h]
   \centering
  \includegraphics[width=0.45\textwidth]{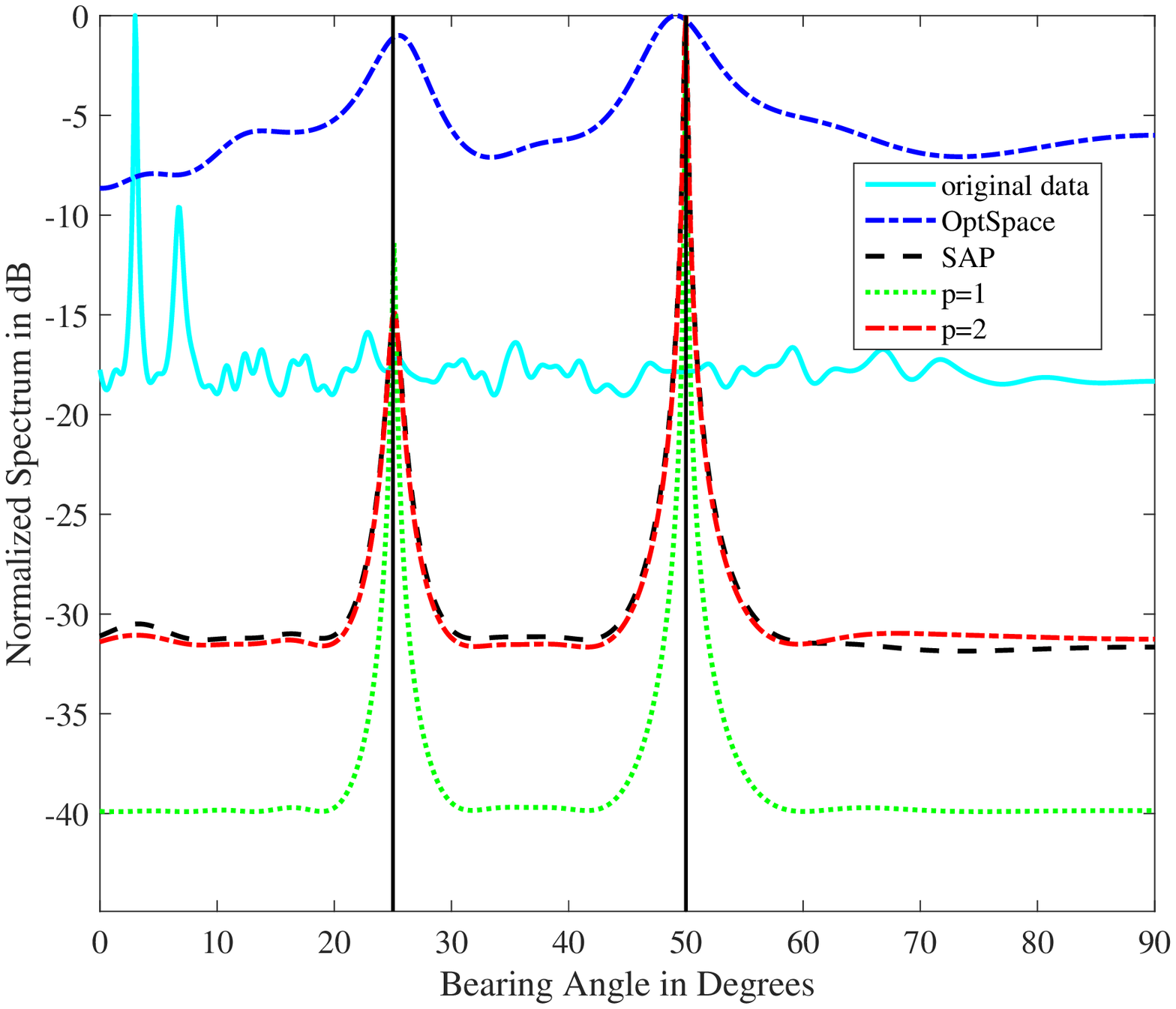} 
  \caption{The comparison of beam pattern.} 
  \label{Fig:5} 
\end{figure}

Fig. \ref{Fig:5} shows the normalized spectrum versus bearing angle in degrees. Due to the effect of sensor failure and reverberation interference on the original data, the normalized spectrum appears false peaks. Hence, the beam pattern is sensitive to the contaminated signal. 
Besides, the main lobe of the OptSpace algorithm converges to the true signal orientation, but there are certain apparent estimation errors.Thirdly, the algorithms of SAP and  $\ell_{2}$-norm have obvious performance improvements,  and both of them can make more accurate spectrum estimation and have almost the same performances. 
Fourthly, the algorithm based on $\ell_{1}$-norm exhibits the narrowest beamwidth and lowest sidelobe compared with other algorithms. Most importantly, the algorithm based on  $\ell_{1}$-norm has the best robustness of spectrum estimation in the ocean reverberation with sensor failure.

\begin{figure*}[h]
\centering
\subfigure[  ]{               
\includegraphics[width=0.3\linewidth]{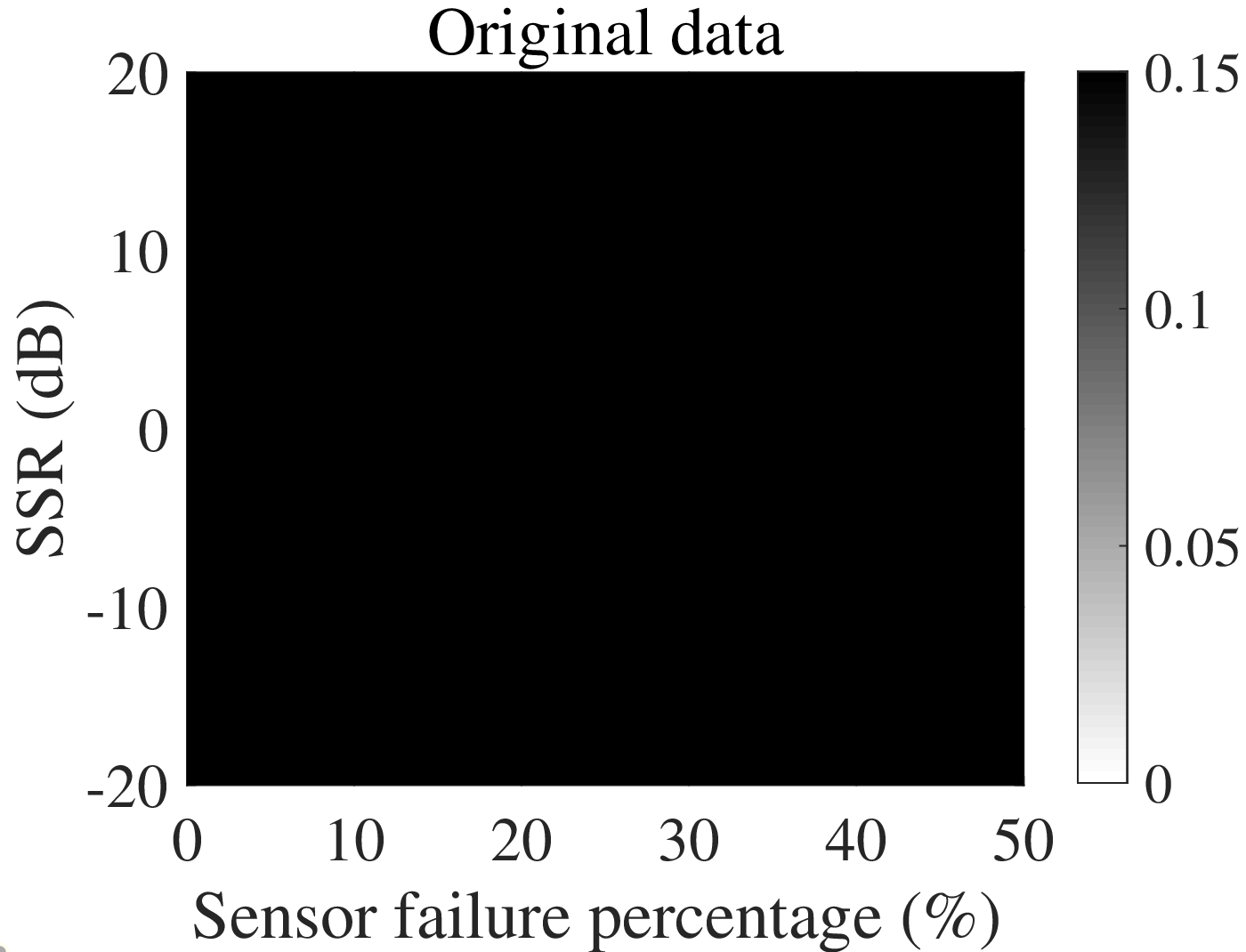}}
\hspace{0mm}
\subfigure[  ]{
\includegraphics[width=0.3\linewidth]{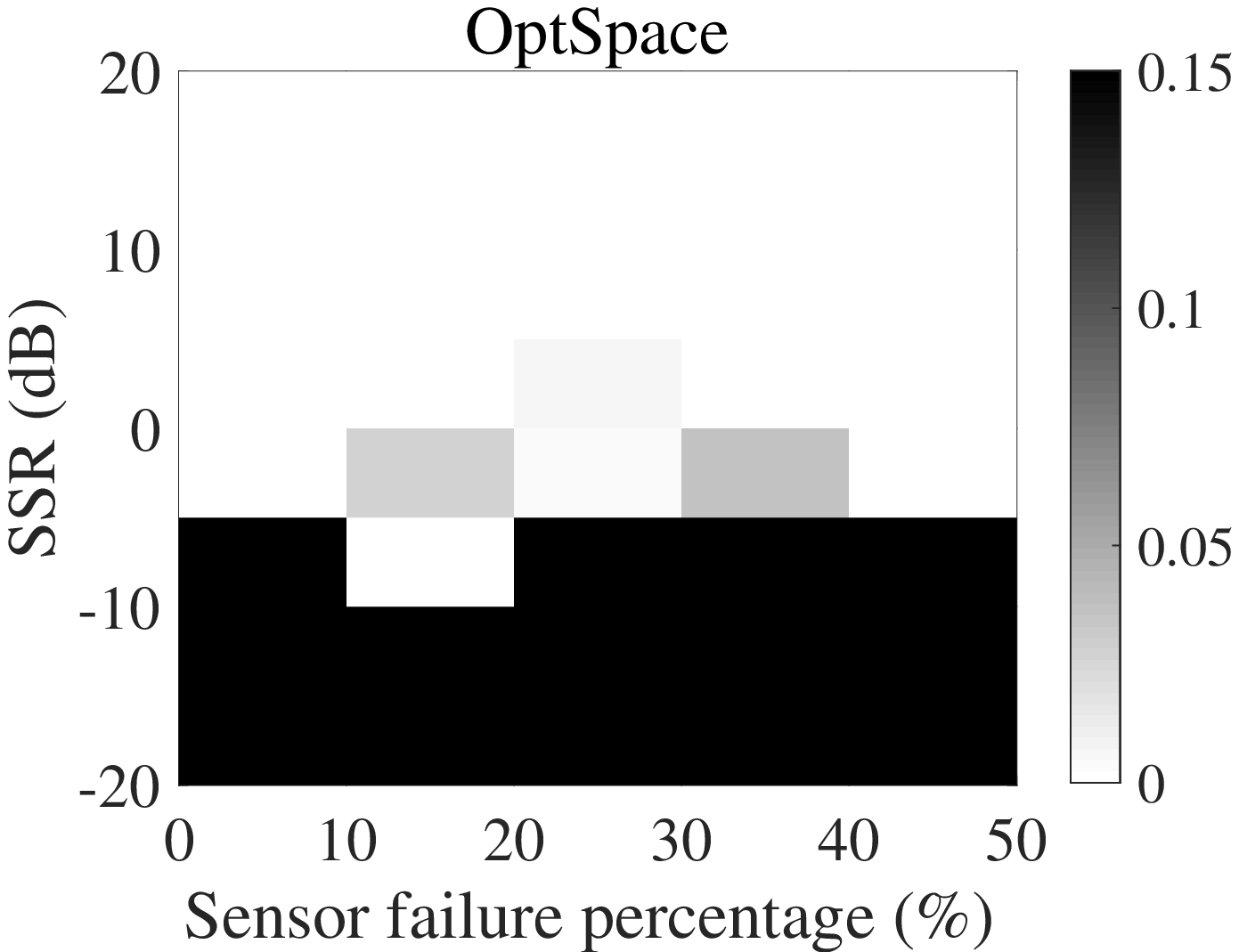}}
\hspace{0mm}
\subfigure[  ]{
\includegraphics[width=0.3\linewidth]{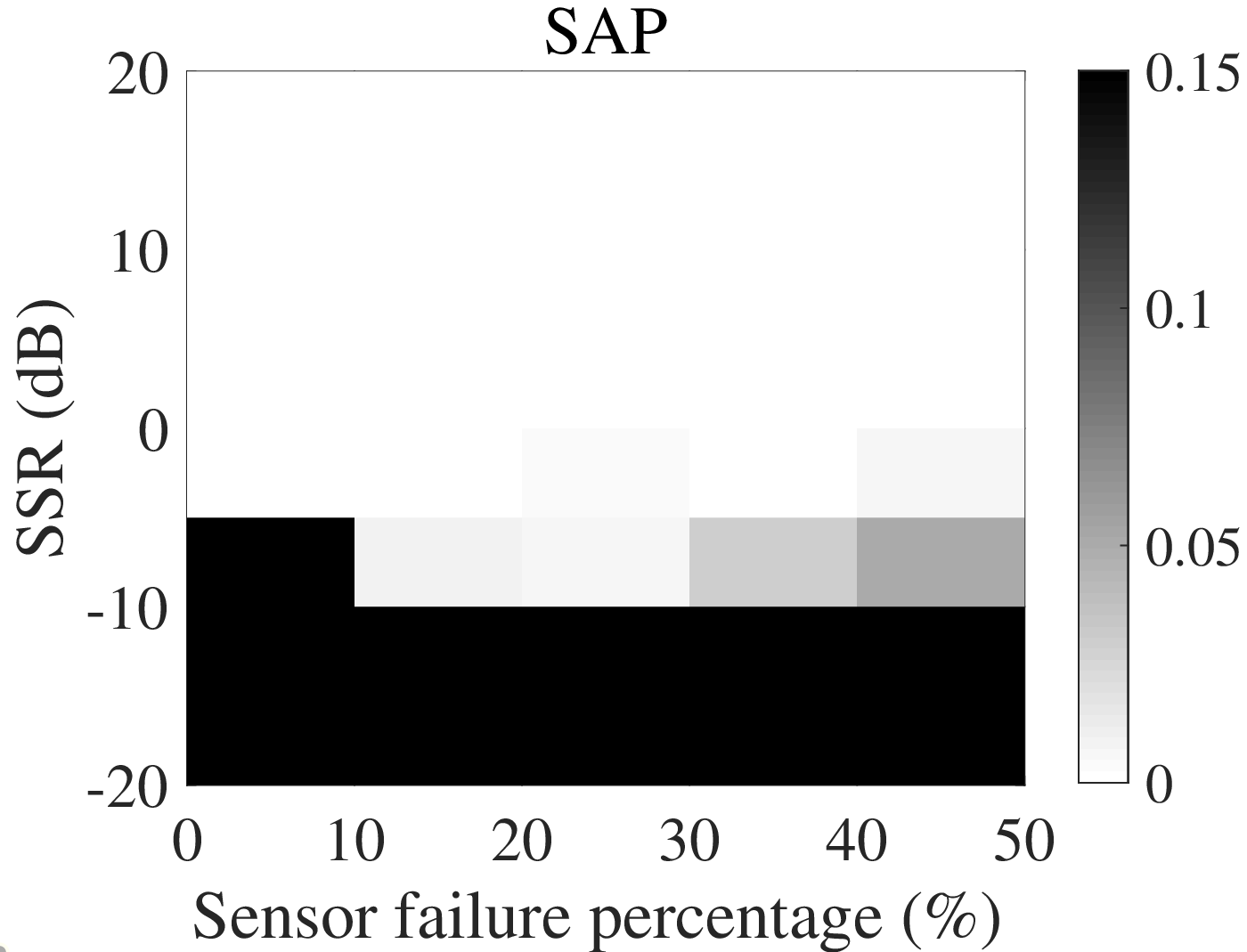}}
\hspace{0mm}
\subfigure[  ]{
\includegraphics[width=0.3\linewidth]{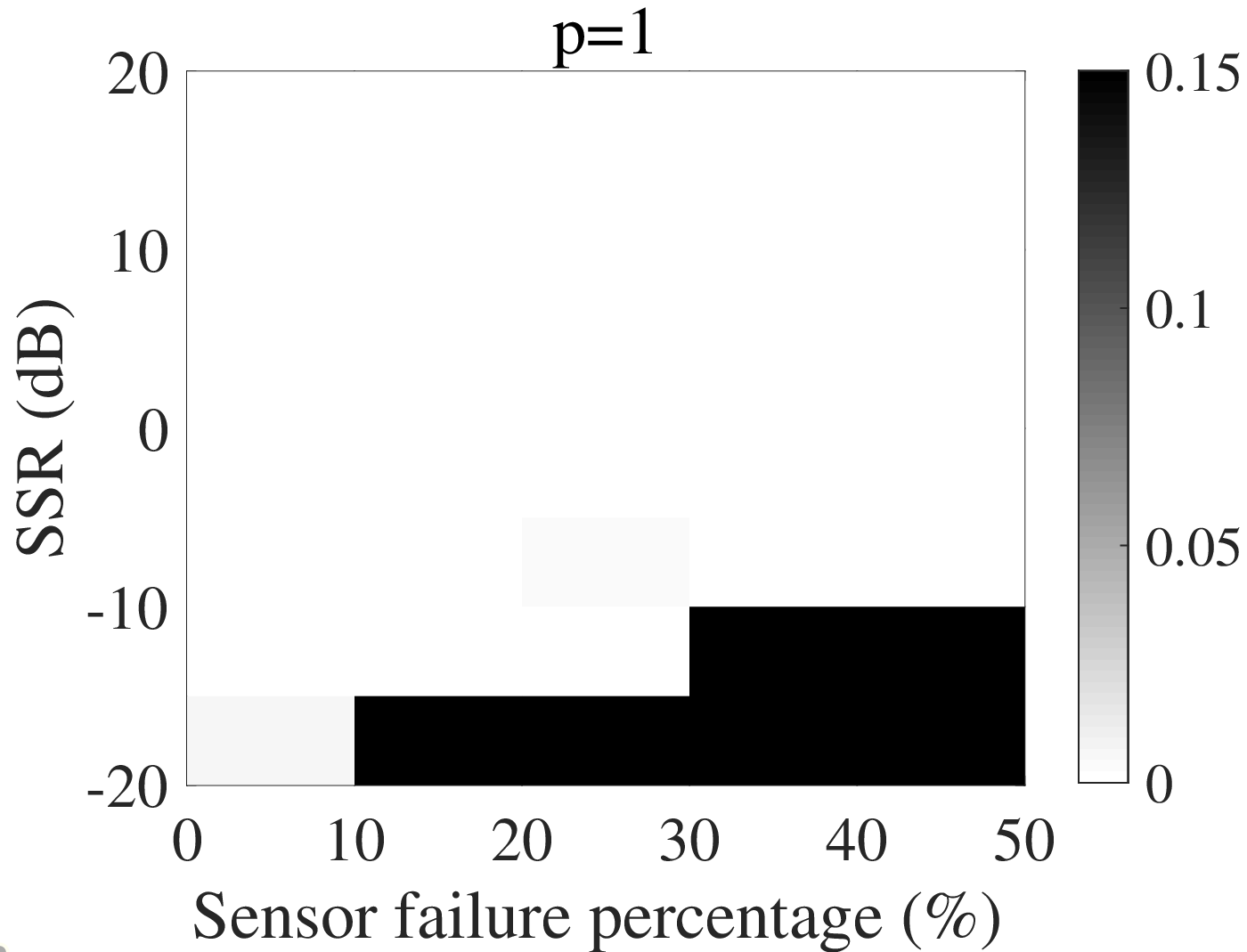}}
\hspace{0mm}
\subfigure[  ]{
\includegraphics[width=0.3\linewidth]{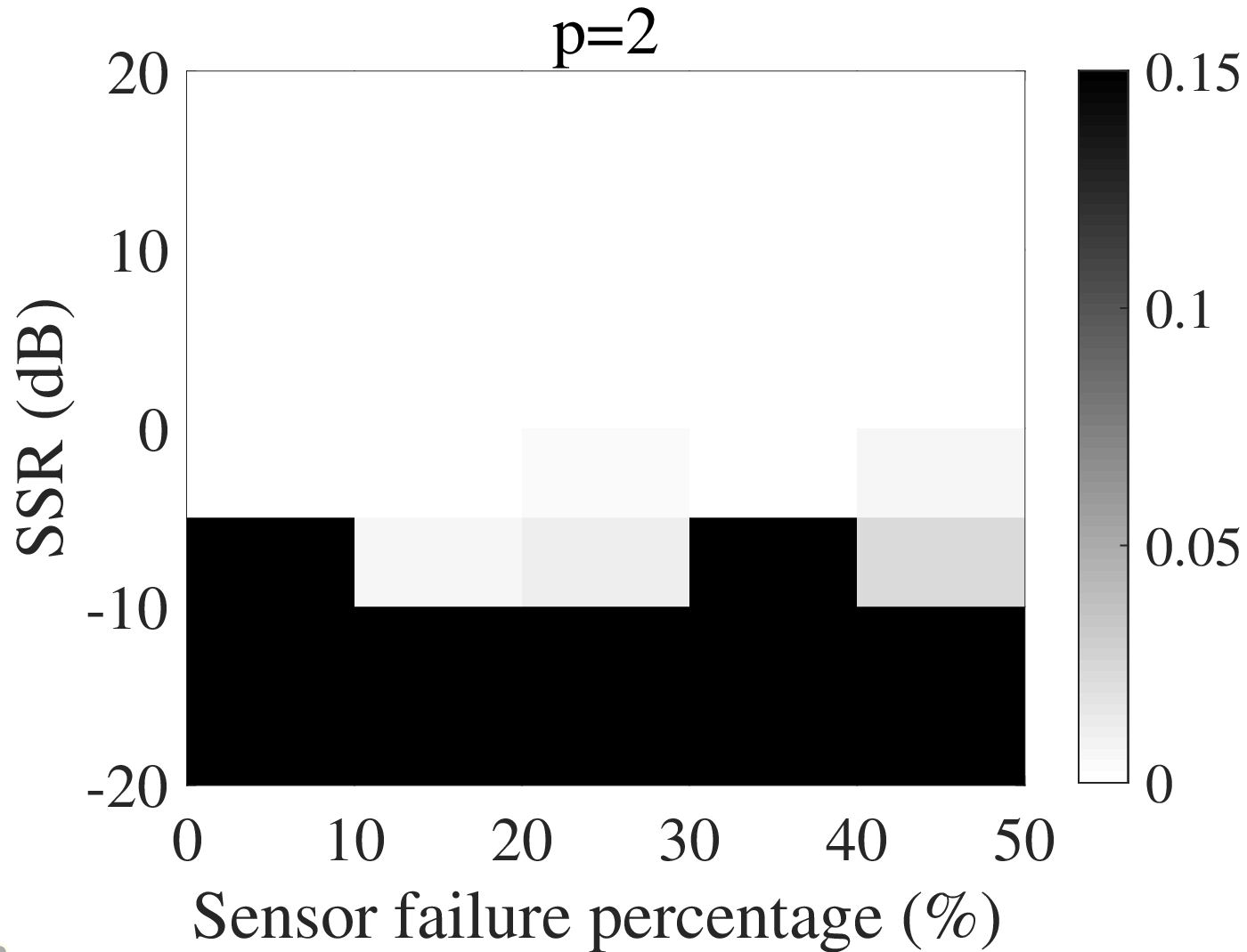}}
\hspace{0mm}
\centering\caption{ RMSE ($\theta$) versus SSR and senor failure percentage.}
\label{Fig:6}
\end{figure*}

According to Fig. \ref{Fig:4}, we choose the sensor failure percentage less than $50\%$ for the comparison of DOA estimation because of the high probability of recovery. Fig. \ref{Fig:6} shows the results of RMSE versus SSR and senor failure percentage, where the threshold of normalized RMSE of DOA is chosen as  0.15. Firstly, if the RMSE is smaller than 0.15, the result can be observed as white region and  dark region otherwise. 
Furthermore, the original data with sensor failure and reverberation interference cause large errors along with different SSRs, in which the all normalized RMSEs definitely exceed 0.15 and show a whole dark region in Fig. \ref{Fig:6} (a). From Fig. \ref{Fig:6} (b)$\sim$(e) it can also be observed that the MC algorithms of OptSpace, SAP and $\ell_{p}$-norm with the receiving acoustic data have a  constructive influence on the DOA estimation. 
Thirdly, when the SSR is above 5 dB, there are all almost no errors of DOA with the involved algorithms. As for the SSR less than 5 dB, the errors of DOA  achieved by the algorithm with $\ell_{1}$-norm will be less. Accordingly, the MC algorithm with $\ell_{1}$-norm is suitable for source location in the ocean reverberation environment with sensor failure, especially for low SSR.

\section{Conclusion}
This paper proposed an approach of acoustic localization in the presence of ocean reverberation with sonar sensor failure. First, the model of sensor failure is established, and the Hankel structured matrix is transformed for the received incomplete matrix. Then, the ocean
reveberation is regarded as outliers in the receiverd matrix. With the low rank factorization, the algorithms of   $\ell_{1}$- and $\ell_{2}$ - norms are  developed to recover the signal matrix in the ocean, which can converge to the ground-truth matrix ultimately. The algorithm based on $\ell_{1}$-norm is suitable for reverberation suppression because of the narrow beam width and low sidelobe. The results show that the robustness of acoustic localization in the ocean reverberation with sensor failure can be improved by the matrix completion with $\ell_{1}$-norm minimization.

%
\ifCLASSOPTIONcaptionsoff
   \newpage
\fi


%

\normalsize
\bibliographystyle{IEEEtran}


\bibliography{reference}
\end{document}